\documentclass[fleqn,usenatbib]{mnras}

\usepackage{newtxtext,newtxmath}

\usepackage[T1]{fontenc}

\DeclareRobustCommand{\VAN}[3]{#2}
\let\VANthebibliography\thebibliography
\def\thebibliography{\DeclareRobustCommand{\VAN}[3]{##3}\VANthebibliography}

\usepackage{graphicx}	

\title[Statistical reliability of Venus 267 GHz JCMT  observations]{The statistical reliability of  267 GHz JCMT  observations of Venus: \newline No significant evidence for phosphine absorption}

\author[]{
M. A. Thompson,$^{1}$\thanks{E-mail: m.a.thompson@herts.ac.uk}
\\
$^{1}$Centre for Astrophysics Research, Department of Physics Astronomy \& Mathematics,\\ University of Hertfordshire, College Lane, Hatfield, AL10 9AB, U.K.
}

\date{Accepted XXX. Received YYY; in original form ZZZ}

\pubyear{2020}

\begin{document}
\label{firstpage}
\pagerange{\pageref{firstpage}--\pageref{lastpage}}
\maketitle

\begin{abstract}
In the light of the recent announcement of the discovery of the potential biosignature phosphine in the atmosphere of Venus I present an independent reanalysis of the original JCMT data to assess the statistical reliability of the detection. Two line detection methods are explored, low order polynomial fits and higher order multiple polynomial fits.  A non-parametric bootstrap analysis reveals that neither line detection method is able to recover a statistically significant detection. Similar to the results of other reanalyses of ALMA Venus spectra, the polynomial fitting process results in false positive detections in the JCMT spectrum. There is thus no significant evidence for phosphine absorption in the JCMT Venus spectra.   
\end{abstract}




\section{Introduction}

Phosphine (PH$_{3}$) is a well-known trace constituent of gas giant atmospheres, first identified within  Jupiter in the mid-infrared by \citet{ridgway1976} and within Saturn in the millimetre and far-infrared by \citet{weisstein1994} and \citet{davis1996}. The main terrestrial sources of PH$_{3}$ are either biogenic (via  anaerobic processes) or anthropogenic (via  agricultural production and semiconductor doping), see further references in \citet{sousa-silva2020} for more details. The biogenic production route and the short chemical lifetime of PH$_{3}$ in an oxidising planetary atmosphere has led to the suggestion that PH$_{3}$ may be a potential biosignature in exoplanet atmospheres \citep{sousa-silva2020}.

Recently, \citet{greaves2020} presented the potential detection of PH$_{3}$ in the atmosphere of Venus via a candidate spectral feature at 267 GHz with the JCMT and ALMA. PH$_{3}$ is not expected to be prevalent in the highly oxidised Venusian atmosphere. If this feature is truly identified with PH$_{3}$, it implies a concentration several orders of magnitude larger than abiogenic Venusian PH$_{3}$ production could sustain \cite{bains2020}. Excluding a wide range of asteroidal delivery and abiogenic production processes \citet{greaves2020} and \citet{bains2020} conclude that an unknown, potentially biogenic, atmospheric process could be at work.

The existence of a Venusian atmospheric biosphere was first proposed by \citet{morowitz1967}. It has been noted by many authors that at an altitude of 50 km, the Venusian atmosphere has a remarkably similar temperature and pressure to that of Earth. In the words of \citet{landis2003}, ``\ldots viewed in a different way, the problem with Venus is merely that the ground level is too far below the one atmosphere level''. If the candidate PH$_{3}$ feature identified by \citet{greaves2020} is confirmed to be real  and all abiogenic production routes can be ruled out this would  lend important support to the \citet{morowitz1967} hypothesis and  be a potential step towards the discovery of a novel non-terrestrial biosphere.

 Since the \citet{greaves2020} claim, there has been much speculation as to the implication and significance of the results \citep[e.g.][]{lingam2020,siraj2020}. \citet{hein2020} have proposed a precursor mission for a balloon-based in-situ search for Venusian life, which could launch as early as 2022.
However, the \citet{greaves2020} results are not without controversy, due to the complex and difficult nature of the observations and assumptions within the chemical network of \citet{bains2020}. 

An independent reanalysis of the ALMA Venus data was performed by \citet{snellen2020} who showed that the analysis technique of \citet{greaves2020} as applied to a spectrum dominated by non-Gaussian noise can result in false positive detections. A more  conservative lower order baseline removal resulted in a candidate feature in the ALMA Venus spectrum of only 2$\sigma$. A further independent reanalysis of the ALMA Venus data has been conducted by \citet{villanueva2020}, who again concluded that the ALMA line detection is a false positive. Atmospheric modelling with a more realistic SO$_{2}$ vertical profile additionally shows that the depth of the JCMT line feature is entirely consistent with the nearby SO$_{2}$ transition \citep{villanueva2020}.  

However, the JCMT Venus spectra from \citet{greaves2020} are also analysed by fitting multiple high order polynomials and the reported JCMT line feature may also be a false positive detection. Motivated by the investigation of \citet{snellen2020}  into the ALMA Venus spectrum  this letter seeks to determine the statistical reliability of the original JCMT single-dish detection and reveal whether this line feature is a significant absorption line or not.

\section{Reanalysis of JCMT data}

\begin{figure}
	\includegraphics[width=\columnwidth]{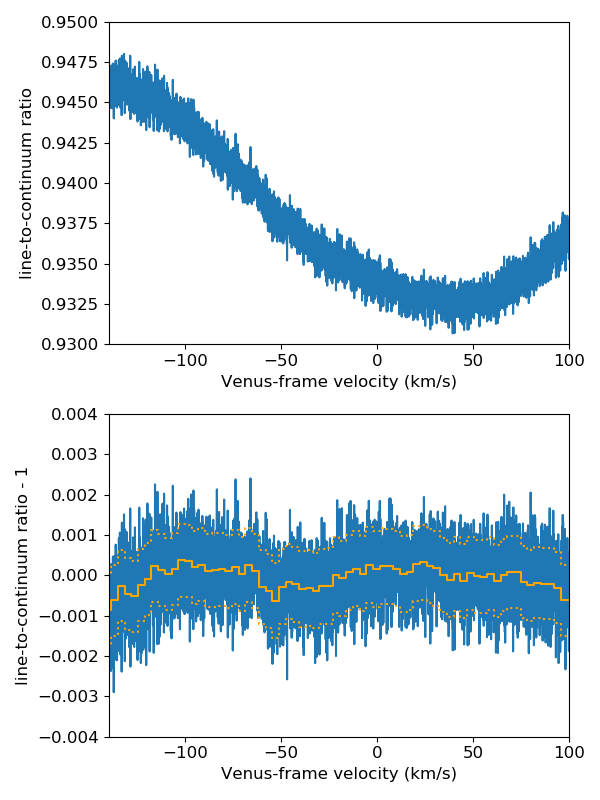}
    \caption{\emph{Top:} Line-to-continuum  spectrum of Venus obtained by integrating the individual time-series spectra and dividing by the mean continuum level of each spectrum. \emph{Bottom:} the above line-to-continuum spectrum with a 3rd-order polynomial subtracted  and denoted as line-to-continuum ratio $-$ 1. The inset lower velocity resolution spectrum is the result of rebinning to 3.5 km\,s$^{-1}$ channels. The dotted orange lines represent the $\pm$1$\sigma$ standard deviation of each  3.5 km\,s$^{-1}$ channel, calculated from the spread of the values from the individual spectra,  revealing substantial dispersion within each channel over time.}
    \label{fig:1}
\end{figure}

The  reduction and analysis procedure for the JCMT Venus data is comprehensively described in \citet{greaves2020}. Here the basic steps are briefly given to illustrate our reanalysis. The aim is to follow as closely as possible the steps carried out by \citet{greaves2020} by modifying the data reduction scripts supplied in that publication. The raw data were obtained from the JCMT Science Archive as described in \citet{greaves2020} and are stored as time series spectral datacubes calibrated on the T$_{A}^{*}$ scale. 512 channels were blanked from the edges of each spectrum to remove noise present at the edge of the passband. The top panel of Figure \ref{fig:1} shows a composite native velocity-resolution (0.0347 km\,s$^{-1}$ channels) spectrum obtained by integrating the Venus spectra along the time axis and dividing by the mean continuum value to obtain a line-to-continuum spectrum. The bottom panel of Figure \ref{fig:1} shows the same spectrum with a 3rd-order polynomial baseline fitted to and subtracted from the entire spectrum, additionally rebinned to 3.5 km$^{-1}$ channels to maintain a common velocity resolution with \citet{greaves2020}.  It must be noted that these baseline-subtracted spectra should strictly be referred to as the ``line-to-continuum ratio $-$ 1'', as they are zero-centred. However to maintain consistency with \citet{greaves2020} we will refer to these spectra as line-to-continuum in the following text.

As Figure \ref{fig:1} shows, the JCMT spectra are  affected by baseline ripples, which are likely to be caused by reflections between the main and cold-load dewar, an unidentified surface in the receiver cabin and the well-known ``JCMT 16-MHz ripple''  caused by a standing wave between the secondary mirror and the receiver cabin\footnote{https://www.eaobservatory.org/jcmt/instrumentation/heterodyne/observing-modes/}. This underlines the extreme difficulty of identifying faint spectral features in the presence of a strong background continuum source.

\subsection{Low order polynomial fits}
\label{sect:low-order}

The simplest way to identify a spectral line in a spectrum with a complex baseline is to fit a low order polynomial to spectral channels bracketing the suspected line position. \citet{greaves2020} carried out such a simple approach as an independent test of their high order polynomial fits by integrating (or collapsing in JCMT parlance) the time-series spectral cubes along the time axis and fitting a low order polynomial. It is reported that the phosphine line was recovered with a lower signal-to-noise-ratio than the higher order fitting approach \citep{greaves2020}. 

The data reduction scripts presented by \citet{greaves2020} were modified to repeat this simple approach and investigate the significance of the line recovery with a single low order polynomial. Polynomials of 3rd and 4th-order were fitted to the spectrum shown in the top panel of Figure \ref{fig:1}. To investigate the effects of different fitting ranges these fits were carried out over varying width spectral regions centred on the PH$_{3}$ Venus rest-frame velocity. A line region of  $|\Delta v| = 5$ km\,s$^{-1}$ centred on the PH$_{3}$ J=1--0 velocity was excluded from the fit, following \citet{greaves2020}. The fitted spectra were then rebinned to a common channel width of 3.5 km\,s$^{-1}$.

\begin{figure}
	\includegraphics[width=\columnwidth]{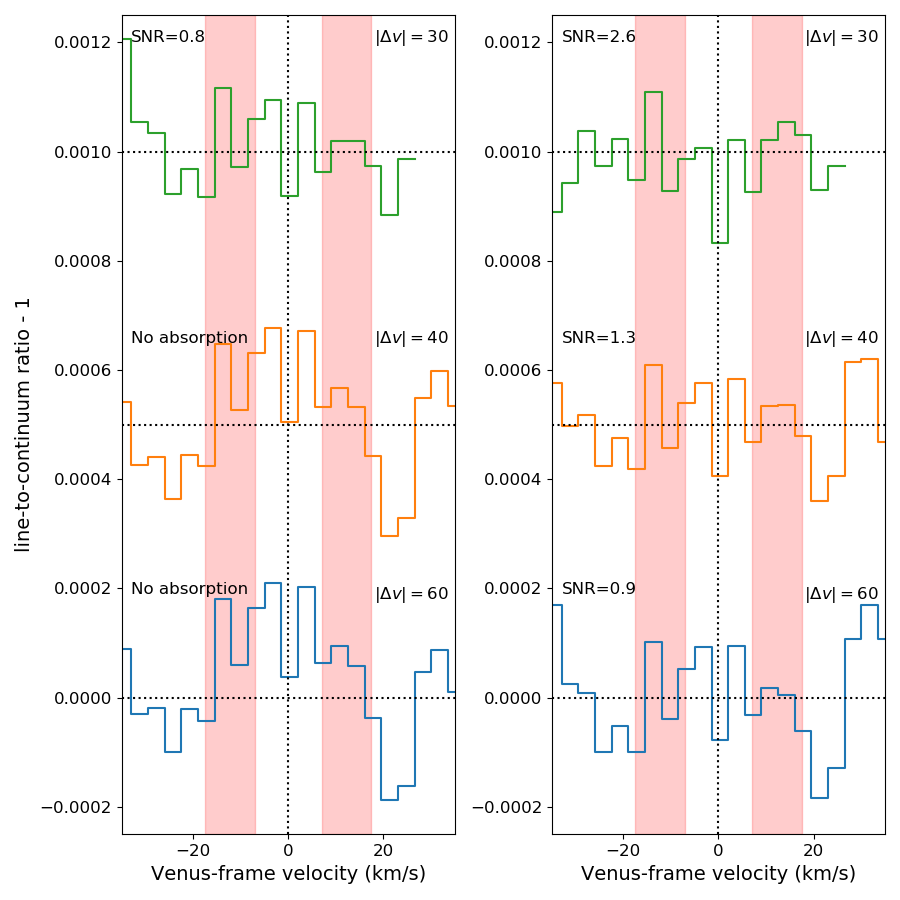}
    \caption{The results of 3rd-order (\emph{left column}) and 4th-order (\emph{right column}) polynomial fits to velocity windows centred on the PH$_{3}$ Venus rest-frame velocity. Spectra are offset vertically for clarity. The width of each velocity window is indicated in the top right of each spectrum and  the signal-to-noise ratio of any identified absorption feature in the central line window is indicated in the top left. Standard deviations were determined from the same line-free channels in each spectrum, as shown by shaded regions. Horizontal and vertical dashed lines indicate the baseline and velocity origins}
    \label{fig:2}
\end{figure}

Figure \ref{fig:2} shows the results of these polynomial fits.  The line search window of  $|\Delta v| = 5$ km\,s$^{-1}$ that was excluded from the polynomial fit is indicated by the central unshaded region in each plot. As can be seen, the result of the fit is sensitive to both the velocity range of channels used in the fitting process and the order of the polynomial used. A low-significance single-channel line feature at the PH$_{3}$ rest frame Venus velocity (with SNR of 2.6) is only recovered with a 4th-order fit over our smallest velocity range ($|\Delta v| \le 30$ km\,s$^{-1}$).

However, the ability of a polynomial fitting routine to incorrectly identify a negative or positive baseline deviation as a spectral line is magnified by giving the routine fewer constraints  and a higher degree of freedom \citep[e.g.~a smaller velocity range and/or a higher order fit;][]{snellen2020}. As also noted by \citet{snellen2020}, the non-Gaussian nature of the spectral noise  makes line identifications more difficult. The usual approach of using the statistical $z-$score (or signal-to-noise ratio) of the peak line temperature becomes meaningless when the noise spectrum does not follow a normal or Gaussian distribution. 

Non-parametric methods are an ideal way to investigate the significance of measurements in a non-Gaussian sample, of which one example is the bootstrap resampling procedure. This process was used to analyse the significance of the recovered line feature in the following way. A Python wrapper was written around the modified \citet{greaves2020} scripts to permit polynomial baseline fits to be repeatedly carried out for different $|\Delta v| = 30$ km\,s$^{-1}$ velocity windows centred on each channel of the native-resolution Venus spectrum. Within each window we identified the minimum value within the central $|\Delta v| = 5$ km\,s$^{-1}$ line region excluded from the fit. 

This approach is analogous to the non-parametric bootstrap method, resampling velocity windows from the original raw spectrum and carrying out the same baseline-fitting and line-search approach to each window. As such, this method is more powerful than the arbitrary sample of velocities presented in Figure 2 of \citet{snellen2020}, as the fraction of realisations that exceed a set peak line-to-continuum ratio can thus be used to estimate the probability that a particular line feature is significant (as a pseudo probability density function or PDF).

It is important to note that the statistical power of this technique is limited due to the fact that there are only  $\sim$4500 possible contiguous velocity windows that can be resampled from the native-resolution  0.0347 km\,s$^{-1}$ spectrum. Greater statistical power could be obtained by relaxing the constraint that each window must be contiguously drawn from the original spectrum, instead filling each window with randomly chosen channels from the entire original spectrum. However this would  randomize any underlying ripples in the baseline structure and it was felt that greater statistical power was not warranted in this case.  

\begin{figure}
	\includegraphics[width=\columnwidth]{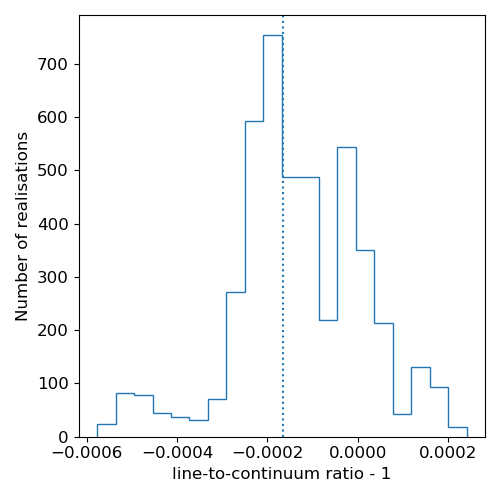}
    \caption{Histogram of the peak line-to-continuum features extracted from the bootstrap process. The vertical dotted line represents the peak line-to-continuum ratio of the line feature detected at the PH$_{3}$ rest velocity in the 4th-order polynomial fit shown in Figure \ref{fig:2} ($-$1.66$\times$10$^{-4}$). }
    \label{fig:3}
\end{figure}

Figure\ref{fig:3} shows a histogram created from our bootstrap realisations, plotting the frequency of the minimum values of line-to-continuum obtained in the line-search region.  The distribution of these values is distinctly non-Gaussian, however it can be seen that the observed depth of the candidate PH$_{3}$ absorption line lies towards the centre of the distribution. In fact  65\% of our bootstrap realisations produce a deeper feature than the candidate PH$_{3}$ absorption line. The null hypothesis that the candidate PH$_{3}$ feature is drawn from a random sample from the spectrum thus cannot be excluded.


\subsection{Higher order polynomial fits}
\label{sect:high-order}

The main analysis procedure used in \citet{greaves2020} to identify the candidate PH$_{3}$ absorption line involves repeated multiple fitting of polynomials to remove a complex spectral baseline. The chosen order of each polynomial are motivated by particular instrumental and observational effects. An initial 4th order baseline is fitted to each observed spectrum to remove a ripple caused by reflections between main and cold load dewars. A further 9th order polynomial is subtracted from a boxcar-smoothed spectrum to remove a reflection from within the receiver cabin. Finally, an 8th order polynomial is subtracted from a  100 km\,s$^{-1}$ wide window between $-36$--$64$ km\,s$^{-1}$ to remove baseline ripples of similar width to the expected line.  

\citet{snellen2020} have shown that 12th-order fits to the ALMA Venus spectrum result in false positive detections at arbitrarily chosen line-free velocities. This effect is caused by the exclusion of the central line region in the polynomial fit. A polynomial of progressively higher order has more freedom to interpolate across the excluded region and thus magnify the significance of spectral features. 

\begin{figure}
	\includegraphics[width=\columnwidth]{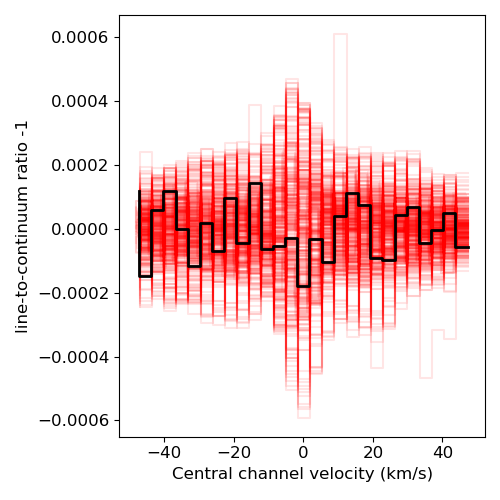}
    \caption{Composite heat-map spectrum of different realisations from the bootstrap resampling process. Only 10\% of realisations are shown for clarity. All spectra have been shifted to a common velocity frame centred on the middle of each 100 km\,s$^{-1}$ window used for the baseline fitting. The spectrum at the PH$_{3}$ Venus rest-frame velocity is indicated by a thick solid black line.}
    \label{fig:4}
\end{figure}

A similar bootstrapping process to that described in Section \ref{sect:low-order} was carried out in order to assess the reliability of the candidate PH$_{3}$ spectral feature presented in \citet{greaves2020} and to test whether the same false positive effects are present in the JCMT data analysis. A Python wrapper script was  used to repeatedly apply the \citet{greaves2020} reduction script to  100 km\,s$^{-1}$ wide windows centred on different central channels drawn from the original spectra. The minimum of each feature within the  $|\Delta v| = 5$ km\,s$^{-1}$ line region excluded from the fit was then stored.

\begin{figure}
	\includegraphics[width=\columnwidth]{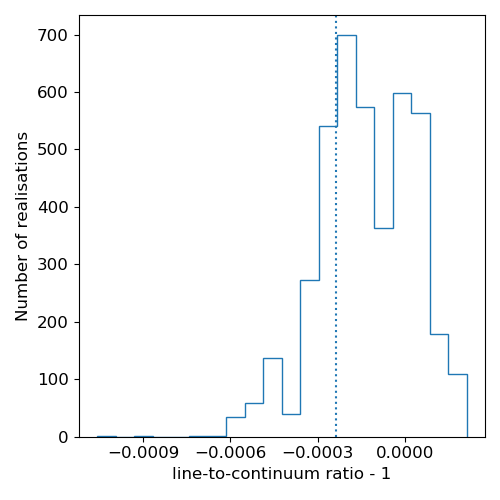}
    \caption{Histogram of the peak line-to-continuum features extracted from the bootstrap resampled multiple polynomial fits described in Section \ref{sect:high-order}. The vertical dotted line represents the peak line-to-continuum ratio of the line feature detected at the PH$_{3}$ rest velocity in the 4th-order polynomial fit shown in Figure \ref{fig:4} (2.38$\times$10$^{-4}$). }
    \label{fig:5}
\end{figure}

The results of this bootstrap analysis are displayed in Figures \ref{fig:4} and \ref{fig:5}. Figure \ref{fig:4} presents a composite heat-map spectrum of a sample of realisations drawn from the bootstrap analysis. Only 10\% of realisations are shown for clarity, drawn uniformly across the distribution of central velocities. The PH$_{3}$ Venus rest-frame velocity realisation is shown as a solid thick black line, and can be seen to be a close match to that in \cite{greaves2020} with minor differences likely to be caused by slightly different channelisations. However, it can also be seen that the depth of the candidate absorption line in this realisation does not appear significant compared to the other realisations. There are many realisations with deeper candidate absorption line features and, indeed, candidate emission line features.

Figure \ref{fig:5} shows the results of all the bootstrap realisations, again as a histogram of the frequency of the minimum values of line-to-continuum obtained in the line-search region. It can be seen that the depth of the candidate PH$_{3}$ feature is towards the centre of the distribution. The peak line-to-continuum ratio of this feature is more significant than that obtained by low order polynomial fitting. However  25\% of our bootstrap realisations produce deeper spectral features than the candidate and the null hypothesis cannot be excluded with any significance. It must be concluded that the candidate line detection presented by \citet{greaves2020} is not statistically significant. 

\section{Conclusions}

An independent reanalysis of the 267 GHz JCMT observations of Venus \citep{greaves2020} is presented. Two different line recovery methods were tested: a simple approach fitting low-order polynomials to line-free channels around the position of the PH$_{3}$ candidate; and the more complex multiple high-order polynomial approach used by \citet{greaves2020}. Similarly to the independent ALMA reanalysis of \citet{snellen2020} and \citet{villanueva2020} it is found that the high-order polynomial fitting applied to the JCMT spectra also results in false positive detections. 

The statistical significance of identified line features was explored by a bootstrap analysis with the result that neither low-order fitting or the high-order fitting method result in statistically significant detections. The null hypothesis that the line candidates are drawn from a random line-to-continuum sample within the spectrum cannot be excluded. This means that there is no significant evidence for either PH$_{3}$ or SO$_{2}$ absorption within this data. 

The investigation carried out in this paper \citep[and also][]{snellen2020,villanueva2020} underlines the difficulty of identifying line features against bright continuum sources and the need for careful and rigorous analysis to exclude false positive detections.

\section*{Acknowledgements}

I would like to thank Chiaki Kobayashi for persuading me to give a Friday lunch talk on this topic, for which preparation eventually resulted in this work. I would also commend the authors of \citet{greaves2020} for making their scripts freely available which considerably aided in the reanalysis of the raw data.  I would like to thank the anonymous referee for their thoughtful and constructive report which has improved this paper.  The James Clerk Maxwell Telescope is operated by the East Asian Observatory on behalf of The National Astronomical Observatory of Japan; Academia Sinica Institute of Astronomy and Astrophysics; the Korea Astronomy and Space Science Institute; Center for Astronomical Mega-Science (as well as the National Key R\&D Program of China with no. 2017YFA0402700). Additional funding support is provided by the Science and Technology Facilities Council of the United Kingdom and participating universities in the United Kingdom and Canada. As JCMT users, we express our deep gratitude to the people of Hawaii for the use of a location on Mauna Kea, a sacred site. This research would not have been possible without the NASA ADS, the Starlink software suite and the astropy package collection. MAT acknowledges support from the UK's Science \& Technology Facilities Council [grant number ST/R000905/1]. 

\section*{Data Availability}

The raw data analysed here was observed under JCMT Service Program S16BP007 and can be obtained from the JCMT Science archive hosted at the Canadian Astronomical Data Centre.



\bibliographystyle{mnras}
\bibliography{venus} 





\bsp	
\label{lastpage}
\end{document}